\documentclass[journal]{IEEEtran}

\usepackage[
    hidelinks,
    pdfborder={0 0 0},
    colorlinks=true,
    linkcolor=blue,
    citecolor=blue,
    urlcolor=blue
]{hyperref}

\usepackage{graphicx}
\usepackage{cite}
\usepackage{amsmath,amssymb,amsfonts}
\usepackage{enumitem}
\usepackage{textcomp}
\usepackage{array}
\usepackage{dsfont}
\usepackage{subcaption}
\usepackage{xcolor}
\usepackage{courier}
\usepackage{url}
\DeclareMathOperator*{\argmin}{\arg\!\min}
\renewcommand{\Re}{\operatorname{Re}}

\begin{document}

\title{SR-ANM: A Square-Root Atomic Norm Minimization Estimator for Noise-Robust Direction-of-Arrival Estimation}

\author{Youval Klioui

\thanks{Youval Klioui is with the Department of Electrical Engineering, Eindhoven University of Technology, Eindhoven 5612 AP, The Netherlands (email: y.klioui@tue.nl)}
}

\hypersetup{
    pdfauthor={},
    pdftitle={},
    pdfsubject={},
    pdfkeywords={}
}

\maketitle

\begin{abstract}
	This paper proposes a Square-root Atomic Norm Minimization (SR-ANM) estimator for direction-of-arrival (DoA) estimation in the single measurement vector (SMV) setting. This new framework, inspired by the Square-root Least Absolute Shrinkage and Selection Operator (Square-root LASSO), relies on a residual term based on the $L_{2}$-norm rather than the usual squared $L_{2}$-norm so as to obtain an estimator whereby the hyperparameter of the atomic norm in the unconstrained convex optimization problem exhibits robustness to the noise level compared to classical ANM. Experimental verification across a wide signal-to-noise ratio (SNR) range from 0 dB to 40 dB using the the probability of detection and the angular root-mean-squared error (RMSE) confirms the increased noise robustness of the proposed framework compared to ANM.
\end{abstract}


\IEEEpeerreviewmaketitle
\section{Introduction}
\IEEEPARstart{A}{tomic} Norm Minimization (ANM) \cite{anm, anm2, ram} is considered a state-of-the-art method for direction-of-arrival (DoA) estimation under a single measurement vector (SMV) setting, as it is able to overcome the grid mismatch effect \cite{mismatch}, one of the most fundamental bottlenecks in grid-based compressed sensing methods such as the $L_{1}$-norm-based Least Absolute Shrinkage and Selection Operator (LASSO) \cite{lasso}, by directly formulating the sparse recovery problem in the continuous atomic set and providing a corresponding semidefinite problem formulation for computing the atomic norm $\|\cdot\|_{\mathcal{A}}$. While able to show a much stronger sparse recovery performance, the ANM framework, similarly to the LASSO framework, nevertheless requires the adjustment of a hyperparameter $\kappa$ to weight the contribution of the atomic norm in the unconstrained convex optimization problem. The optimal $\kappa$ typically depends on the noise level $\sigma$ \cite{anm}, a quantity that is difficult to estimate from a single measurement vector alone. Additionally, for a sparse linear array with $M$ elements subsampled from a uniform linear array (ULA) of $\tilde{M}$ elements, the computational complexity per iteration for solving the ANM semidefinite program (SDP) using the alternating direction method of multipliers (ADMM) algorithm is at least $\mathcal{O}(\tilde{M}^{3})$. This makes cross-validation for experimentally determining the optimal $\kappa$ for each measurement vector $\mathbf{y} \in \mathbb{C}^{M}$ prohibitively expensive. Thus, an atomic norm framework that exhibits a reduced sensitivity to the noise level for the adjustment of its sparsity hyperparameter is of practical interest. In order to address this same limitation for the LASSO framework, Belloni et al. introduced the Square-root LASSO framework \cite{sqlasso}, wherein they show that by using an $L_{2}$-norm based residual instead of the classical squared $L_{2}$-norm, the hyperparameter $\kappa$ required for adjusting the strength of the $L_{1}$-norm in the unconstrained objective exhibits a much weaker dependence on the noise level compared to the classical LASSO. The work in this paper provides a similar extension for the ANM framework and the contributions can be summarized as follows:
\begin{itemize}
\item We propose a Square-root Atomic Norm Minimization (SR-ANM) estimator based on an $L_{2}$-norm residual that exhibits an improved robustness to the noise level compared to ANM. We additionally provide a theoretical characterization of the optimality conditions that the global solution to the SR-ANM convex problem must satisfy.
 \item We provide a performance characterization of SR-ANM across a signal-to-noise ratio (SNR) range from 0 dB up to 40 dB using the detection rate ($P_{d}$) and angular root-mean-squared error (RMSE) and experimentally confirm the  noise robustness of the proposed framework compared to ANM.
\end{itemize}
\par Section \ref{sec:anm_and_square_root_lasso} sets up the signal model and provides a review of the ANM and Square-root LASSO frameworks. Section \ref{sec:square_root_anm} introduces SR-ANM and provides a theoretical characterization of its global solution. We additionally provide simple closed-form ADMM updates for solving the SR-ANM convex optimization problem. Section \ref{sec:experimental_results} provides a performance characterization of SR-ANM to confirm the increased noise robustness of the proposed framework compared to ANM. Section \ref{sec:conclusion} concludes the paper.\footnote{The repository for replicating the results reported here can be found at: {\fontfamily{pcr}\selectfont\url{https://github.com/youvalklioui/sr-anm}}} 
\par \emph{\underline{Notation:}} $\mathbb{C}$ and  $\mathbb{N}$ denote the set of complex and natural numbers, respectively. $\mathcal{T}^{H}$ denotes the set of Toeplitz-Hermitian matrices. $|z|$ denotes the magnitude  of the complex number $z$. $(.)^{T}$ and $(.)^{H}$ denote the transpose and Hermitian transpose operators, respectively. $\mathbf{I}_{N}$ denotes the size $N$ identity matrix. The notation $\mathbf{X}\succeq 0$ indicates $\mathbf{X}$ is a positive semidefinite matrix. $\mathbf{X}_{m,.}$ and $\mathbf{X}_{.,n}$ denote the $m$-th row and $n$-th column, respectively, of $\mathbf{X}\in \mathbb{C}^{M\times N}$. $\|\cdot\|_{F}$ and $\|\cdot\|_{\infty}$ denote the Frobenius and maximum norms, respectively. $\langle \mathbf{x}, \mathbf{y}\rangle = \Re(\mathbf{x}^{H}\mathbf{y})$ denotes the real inner product. $\mathds{1}(\mathcal{B})$ is the indicator function with $\mathds{1}(\mathcal{B})=1$ if the statement $\mathcal{B}$ is true, and 0 otherwise. $\partial$ denotes the subdifferential operator.

\section{ANM and Square-root LASSO}
\label{sec:anm_and_square_root_lasso}
We consider a general $M$-element linear array subsampled from an $\tilde{M}$-element ULA with $\lambda/2$ inter-element spacing, with $M<\tilde{M}$, along with $K$ sources located at angular positions $\{\theta_{k}^{*}\}_{k=1}^{K}$  as per the following model
\begin{align}
	\mathbf{y}= \sum_{k=1}^{K}c_{k} \mathbf{\Psi}_{\mathcal{M}}\mathbf{a}(\theta_{k}^{*})+\sigma\mathbf{z},
	\label{signal_model}
\end{align}
where $\mathbf{y} \in \mathbb{C}^{M}$ is the single snapshot measurement vector, $\mathbf{z} \sim \mathcal{CN} (\mathbf{0}, \mathbf{I}_{M})$ is a standard complex random Gaussian vector, $\sigma>0$,  $c_{k} \in \mathbb{C}$ is the complex amplitude of the $k$-th source located at angle $\theta_{k}^{*}$, and where $\mathbf{a}(\theta_{k}^{*}) \in \mathbb{C}^{\tilde{M}}$ is the ULA array steering vector given by,
\begin{align}
	(\mathbf{a}(\theta_{k}^{*}))_{m}=  \exp{\big(-j2\pi \frac{\sin(\theta_{k}^{*})}{2}m\big)}, \qquad m=1, 2, \ldots, \tilde{M},
\end{align}
and $\mathbf{\Psi}_{\mathcal{M}} \in \{0,1\}^{M\times \tilde{M}}$ is a subsampling matrix that selects $M$ elements from the $\tilde{M}$-element ULA. In what follows, we denote by $\mathcal{M} =\{1, m_{2}, \ldots, m_{M-1}, \tilde{M}\}$ the set of linear indices of the subsampled array along with the notation $\mathbf{v}_{\mathcal{M}} = \mathbf{\Psi}_{\mathcal{M}}\mathbf{v}$.  DoA estimation within the LASSO framework can be formulated by defining a discrete angular grid $\{\theta_{n}\}_{n=1}^{N}$ along with a so-called dictionary defined as $\mathbf{A}=\begin{bmatrix} \mathbf{a}(\theta_{1}),  \mathbf{a}(\theta_{2}), \ldots, \mathbf{a}(\theta_{N})\end{bmatrix}$
and solving for a given hyperparameter $\kappa>0$ the following unconstrained convex optimization problem
\begin{align}
\hat{\mathbf{c}} = \argmin_{\mathbf{x}} {\frac{1}{2}\|\mathbf{y}-\mathbf{\Psi}_{\mathcal{M}}\mathbf{A}\mathbf{x}\|_{2}^{2} + \kappa\|\mathbf{x}\|_{1}},
\label{lasso}
\end{align}
where $\hat{\mathbf{c}} \in \mathbb{C}^{N}$ is a sparse vector whose support corresponds to estimates of the true angles $\{\theta_{k}^{*}\}_{k=1}^{K}$. The value of $\kappa$ influences how much weight is given to the $L_{1}$-norm in the objective function, and hence how sparse the solution $\hat{\mathbf{c}}$ is. The optimal value of $\kappa$ is known to depend directly on the noise level $\sigma$. This can be seen from the optimality condition that the solution $\hat{\mathbf{c}}$ to the convex problem in \eqref{lasso} must satisfy, since we have
\begin{align}
\mathbf{0} \in  \left. \partial \left( \frac{1}{2} \Vert{}\mathbf{y} - \mathbf{\Psi}_{\mathcal{M}}\mathbf{A}\mathbf{x}\Vert{}_2^2 + \kappa \Vert{}\mathbf{x}\Vert{}_1 \right) \right\vert{}_{\mathbf{x} = \hat{\mathbf{c}}},
\end{align}
which implies the following set of inequalities 
\begin{align}
|((\mathbf{\Psi}_{\mathcal{M}}\mathbf{A})^{H})_{n,.}(\mathbf{y}-\mathbf{\Psi}_{\mathcal{M}}\mathbf{A}\hat{\mathbf{c}})|  &\le  \kappa, \quad n = 1,\ldots, N \label{kappa_inequality}.
\end{align}
Under a pure-noise observation given by $\mathbf{y} = \sigma\mathbf{z}$, i.e. $c_{k} = 0$ for $k=1,\ldots, K$ in \eqref{signal_model}, the set of inequalities in \eqref{kappa_inequality} for the optimal solution $\hat{\mathbf{c}} = \mathbf{0}$ reduces to 
\begin{align}
 \kappa \ge \sigma\|(\mathbf{\Psi}_{\mathcal{M}}\mathbf{A})^{H}\mathbf{z}\|_{\infty},
\end{align}
where the quantity $\|(\mathbf{\Psi}_{\mathcal{M}}\mathbf{A})^{H}\mathbf{z}\|_{\infty}$ is independent of the noise level $\sigma$. Thus, under the pure noise setting, the minimum threshold required by the hyperparameter $\kappa$ depends linearly on $\sigma$. A treatment of the more general setting can be found in \cite{variance_lasso} wherein the authors show the same linear dependence of $\kappa$ with respect to $\sigma$ under some assumptions for the so-called design matrix $\mathbf{\Psi}_{\mathcal{M}}\mathbf{A}$.

\par Correctly setting the optimal $\kappa$ requires prior knowledge of the noise standard deviation $\sigma$, which is difficult to reliably estimate from a single measurement vector $\mathbf{y}$. To address this shortcoming, Belloni et al. introduced the Square-root LASSO \cite{sqlasso} framework whereby we solve the following optimization problem
\begin{align}
\hat{\mathbf{c}} = \argmin_{\mathbf{x}} {\|\mathbf{y}-\mathbf{\Psi}_{\mathcal{M}}\mathbf{A}\mathbf{x}\|_{2} + \kappa\|\mathbf{x}\|_{1}}.
\label{sqlasso}
\end{align}
The authors showed that the use of the $L_{2}$-norm for the residual as opposed to the squared $L_{2}$-norm has the desirable effect of making $\kappa$ independent of $\sigma$. The problem in \eqref{sqlasso} is still convex, and the optimality condition is therefore given by 
\begin{align}
\mathbf{0} \in  \left. \partial \big(\|\mathbf{y}-\mathbf{\Psi}_{\mathcal{M}}\mathbf{A}\mathbf{x}\|_{2} +\kappa\|\mathbf{x}\|_{1} \big) \right\vert{}_{\mathbf{x} = \hat{\mathbf{c}}},
\label{optimality_sqlasso}
\end{align}
which results in the following set of constraints
\begin{align}
\frac{|((\mathbf{\Psi}_{\mathcal{M}}\mathbf{A})^{H})_{n,.}(\mathbf{y}-\mathbf{\Psi}_{\mathcal{M}}\mathbf{A}\hat{\mathbf{c}})|}{\|\mathbf{y}-\mathbf{\Psi}_{\mathcal{M}}\mathbf{A}\hat{\mathbf{c}}\|_{2}}  &\le  \kappa, \quad n = 1,\ldots, N.
\end{align}
Under the same noise model $\mathbf{y} = \sigma\mathbf{z}$, the optimality condition for a null solution $\hat{\mathbf{c}} = \mathbf{0}$ is given by
\begin{align}
\frac{\sigma |((\mathbf{\Psi}_{\mathcal{M}}\mathbf{A})^{H})_{n,.}\mathbf{z}|}{\sigma\|\mathbf{z}\|_{2}}  \le \kappa, \quad n = 1,\ldots, N,
\end{align}
which reduces to 
\begin{align}
\kappa \ge \left\|\frac{(\mathbf{\Psi}_{\mathcal{M}}\mathbf{A})^{H}\mathbf{z}}{\|\mathbf{z}\|_{2}} \right\|_{\infty}  , 
\label{threshold_sqlasso}
\end{align}
where the right-hand side is independent of $\sigma$ and depends primarily on the design matrix $\mathbf{\Psi}_{\mathcal{M}}\mathbf{A}$ and the realization of the standard complex random Gaussian vector $\mathbf{z}$. In other words, under the Square-root LASSO framework and a purely noisy observation, the minimum value for the hyperparameter $\kappa$ required to obtain the optimal null solution is independent of the noise level $\sigma$. We refer the reader to \cite{sqlasso} for a complete treatment of the more general setting where the authors establish the same independence of $\kappa$ with respect to $\sigma$.

The ANM framework \cite{anm} introduced by Bhaskar et al. can be seen as the continuous extension of the LASSO framework, whereby the grid mismatch effect inherent to grid-based methods is entirely bypassed through the use of a so-called continuous atomic set $\mathcal{A} = \big\{\mathbf{a}(\theta) ~|~  \theta \in[-\pi/2, \pi/2)\big\}$
where 
\begin{align}
\begin{cases}
\mathbf{a}(\theta) &= [e^{j2\pi f(\theta)}, \ldots, e^{j2\pi f(\theta) \tilde{M}}]^{T} \\
f(\theta) &= -\frac{1}{2} \sin(\theta),
\end{cases}
\end{align}
along with the convex atomic norm $\|\cdot\|_{\mathcal{A}}$ defined as \cite{anm_def}
\begin{align}
\|\mathbf{x}\|_{\mathcal{A}}
=
\inf_{\theta_{k}, c_{k}}
\left\{
\sum_{k} |c_k|
\;\middle|\;
\mathbf{x}=\sum_{k} c_k \mathbf{a}(\theta_k)
\right\}.
\end{align}
The ANM optimization problem for the sparse subsampled array is then given by
\begin{align}
\min_{\mathbf{x}}\frac{1}{2}\| \mathbf{y} - \mathbf{\Psi}_{\mathcal{M}}\mathbf{x}\|_{2}^{2} + \kappa \|\mathbf{x}\|_{\mathcal{A}},
\label{anm_obj}
\end{align}
where $\kappa>0$ controls the strength of the atomic norm regularizer. It is worth noting here that, unlike the LASSO, the object being reconstructed in the objective in \eqref{anm_obj} is the denoised measurement vector rather than the sparse representation itself. The atomic norm can be equivalently formulated as the solution to the semidefinite program given by \cite{anm}
\begin{align}
\|\mathbf{x}\|_{\mathcal A} = \inf_{\mathbf{T}\in \mathcal{T}^{H},t} \left\{ \frac12\left(t+\frac{\operatorname{tr}(\mathbf{T})}{\tilde{M}}\right) : \ \begin{bmatrix}\mathbf{T}  & \mathbf{x}\\ \mathbf{x}^H & t\end{bmatrix}\succeq 0 \right\},
\end{align}
where $\mathbf{T}$ is a size $\tilde{M}$ Toeplitz-Hermitian matrix. The SDP formulation of the ANM framework then amounts to solving the following convex problem given the single measurement vector $\mathbf{y}$ from the $M$-element sparse array
\begin{align}
&\min_{\mathbf{T}, \mathbf{x}, t} \frac{1}{2}\| \mathbf{y} - \mathbf{\Psi}_{\mathcal{M}}\mathbf{x}\|_{2}^{2} + \frac{\kappa}{2} \left(t+\frac{\operatorname{tr}(\mathbf{T})}{\tilde{M}}\right), \nonumber \\
	& s.t.  \quad \begin{bmatrix}\mathbf{T}  & \mathbf{x}\\ \mathbf{x}^H & t\end{bmatrix}\succeq 0, \quad \mathbf{T} \in \mathcal{T}^{H}.
\label{anm_obj_sdp}
\end{align}
Given a solution $(\hat{\mathbf{T}}, \hat{\mathbf{x}}, \hat{t})$ to \eqref{anm_obj_sdp}, estimates $\{\hat{\theta}_{k}\}_{k=1}^{K}$ of the angles $\{\theta^{*}_{k}\}_{k=1}^{K}$ can be found by performing a Root-MUSIC \cite{rmusic}\cite{rmusic2} estimation on $\hat{\mathbf{T}}$, which effectively represents a denoised covariance matrix. The hyperparameter $\kappa$ in $\eqref{anm_obj_sdp}$, similarly to LASSO, depends linearly on $\sigma$. Specifically, it can be shown \cite{anm} that for an array of $M$-elements sampled from a ULA of $\tilde{M}$-elements, the ANM estimator is asymptotically efficient for $\kappa = \sigma(M\log\tilde{M})^{1/2}$.
\section{Square-Root ANM}
\label{sec:square_root_anm}
Similarly to the Square-root LASSO, we propose a Square-root ANM (SR-ANM) framework where we use an $L_{2}$-norm residual and solve the following convex problem
\begin{align}
\min_{\mathbf{x}}\| \mathbf{y} - \mathbf{\Psi}_{\mathcal{M}}\mathbf{x}\|_{2}+ \kappa \|\mathbf{x}\|_{\mathcal{A}},
\label{sranm_obj}
\end{align} with $\kappa >0$. We have the following three propositions related to this framework.\\
\emph{Proposition 1:} The two optimality conditions that a solution $\hat{\mathbf{x}}$ to the SR-ANM problem in \eqref{sranm_obj} must satisfy are given by
\begin{enumerate}[label=\roman*)] 
\item $\left\| \frac{\mathbf{\Psi}_{\mathcal{M}}^{T} (\mathbf{y}-\mathbf{\Psi}_{\mathcal{M}}\hat{\mathbf{x}})}{\|\mathbf{y}-\mathbf{\Psi}_{\mathcal{M}}\hat{\mathbf{x}}\|_{2}} \right\|_{\mathcal{A}}^{*} \le \kappa \quad \text{ii)} \quad \langle \frac{\mathbf{\Psi}_{\mathcal{M}}^{T} (\mathbf{y}-\mathbf{\Psi}_{\mathcal{M}}\hat{\mathbf{x}})}{\|\mathbf{y}-\mathbf{\Psi}_{\mathcal{M}}\hat{\mathbf{x}}\|_{2}}, \hat{\mathbf{x}} \rangle = \kappa \|\hat{\mathbf{x}}\|_{\mathcal{A}}$,
\end{enumerate}
where $\| \cdot \|_{\mathcal{A}}^{*}$ denotes the dual norm of the atomic norm \cite{anm} and $\mathbf{y}-\mathbf{\Psi}_{\mathcal{M}}\hat{\mathbf{x}} \neq \mathbf{0}$. \\
\emph{Proof:} The problem in \eqref{sranm_obj} is convex, from which at the solution $\mathbf{x}=\hat{\mathbf{x}}$ we have
\begin{align}
\mathbf{0} \in \partial \left.\left( \| \mathbf{y} - \mathbf{\Psi}_{\mathcal{M}}\mathbf{x} \|_{2} + \kappa \|\mathbf{x}\|_{\mathcal{A}} \right)\right|_{\mathbf{x} = \hat{\mathbf{x}}},
\label{sranm_opt1}
\end{align}
or equivalently,
\begin{align}
\frac{\mathbf{\Psi}_{\mathcal{M}}^{T} (\mathbf{y}-\mathbf{\Psi}_{\mathcal{M}}\hat{\mathbf{x}})}{\kappa\|\mathbf{y}-\mathbf{\Psi}_{\mathcal{M}}\hat{\mathbf{x}}\|_{2}} \in \partial\|\hat{\mathbf{x}}\|_{\mathcal{A}}.
\label{sranm_opt2}
\end{align}
The subdifferential $\partial\|\hat{\mathbf{x}}\|_{\mathcal{A}}$ of the atomic norm is additionally given by \cite{subdiff_anm}
\begin{align}
\partial \|\hat{\mathbf{x}}\|_{\mathcal{A}} = 
\{\mathbf{q} \in \mathbb{C}^{\tilde{M}}: \|\mathbf{q}\|_{\mathcal{A}}^{*} \le 1 ~ \textrm{and} ~ \langle \mathbf{q}, \hat{\mathbf{x}}\rangle = \|\hat{\mathbf{x}}\|_{\mathcal{A}}\}.
\label{subdiff_anm}
\end{align}
Combining \eqref{sranm_opt2} with \eqref{subdiff_anm} and multiplying both sides by $\kappa$ yields the desired optimality conditions. $\square$\\
\emph{Proposition 2:} Under a pure-noise observation $\mathbf{y} = \sigma\mathbf{n}$, with $\mathbf{n} \sim \mathcal{D}$, $P(\mathbf{n}=\mathbf{0})=0$, and $\mathcal{D}$ is independent of $\sigma>0$, the minimum value of $\kappa$ in \eqref{sranm_obj} for obtaining the optimal null solution $\hat{\mathbf{x}} = \mathbf{0}$ is independent of $\sigma$. \\
\emph{Proof:} Substituting $(\mathbf{y}, \hat{\mathbf{x}})=(\sigma\mathbf{n}, \mathbf{0})$ in the optimality conditions, we obtain
\begin{align}
\kappa \ge \left\| \frac{\mathbf{\Psi}_{\mathcal{M}}^{T} (\sigma \mathbf{n})}{\|\sigma \mathbf{n}\|_{2}} \right\|_{\mathcal{A}}^{*} \implies \kappa \ge \left\| \frac{\mathbf{\Psi}_{\mathcal{M}}^{T}\mathbf{n}}{\| \mathbf{n}\|_{2}} \right\|_{\mathcal{A}}^{*},
\end{align}
where the right-hand side is independent of $\sigma$ and depends instead on the geometry of the atomic set $\mathcal{A}$, the subsampling matrix $\mathbf{\Psi}_{\mathcal{M}}$, and the specific realization of $\mathbf{n}$, which is in turn determined by the distribution $\mathcal{D}$ of $\mathbf{n}$. In contrast, for ANM, using a similar derivation, the minimum value of $\kappa$ is $\sigma$-dependent and given by
\begin{align}
\kappa \ge \sigma \left \|\mathbf{\Psi}_{\mathcal{M}}^{T}\mathbf{n} \right\|_{\mathcal{A}}^{*}.\square
\end{align}
It is worth noting that the minimum threshold for $\kappa$ in \cite{anm} that guarantees an asymptotic performance for the ANM estimator is derived from an upper bound on $\left \|\mathbf{\Psi}_{\mathcal{M}}^{T}\mathbf{n} \right\|_{\mathcal{A}}^{*}$, which under a Gaussian distribution $\mathbf{n}\sim \mathcal{CN}(\mathbf{0},\mathbf{I}_{M})$ can be shown to satisfy $\left \|\mathbf{\Psi}_{\mathcal{M}}^{T}\mathbf{n} \right\|_{\mathcal{A}}^{*} \lesssim (M\log\tilde{M})^{1/2}$. Using the approximation $\| \mathbf{n} \|_{2} \approx \sqrt{M}$, it follows that $\left\| \frac{\mathbf{\Psi}_{\mathcal{M}}^{T}\mathbf{n}}{\| \mathbf{n}\|_{2}} \right\|_{\mathcal{A}}^{*} \lesssim (\log\tilde{M})^{1/2}$. We will subsequently show in the experimental section that such a regularizer allows SR-ANM to exhibit a similar asymptotic performance to that of ANM.

\emph{Proposition 3:} For a given measurement $\mathbf{y}\neq \mathbf{0}$, we define the critical value of the hyperparameter $\kappa$ as $\kappa_{max}(\mathbf{y}) = \inf\{{\kappa>0}: \hat{\mathbf{x}}(\mathbf{y}, \kappa) = \mathbf{0}\}$. Then we have 
\begin{align}
\kappa_{max}^{\textrm{SR-ANM}}(\mathbf{y})=\frac{\kappa_{max}^{\textrm{ANM}}(\mathbf{y})}{\|\mathbf{y}\|_{2}}.
\end{align}
\emph{Proof:} Setting $\hat{\mathbf{x}}= \mathbf{0}$ for a given $\mathbf{y}\neq \mathbf{0}$ in the optimality conditions of SR-ANM, we obtain 
\begin{align}
\left\| \frac{\mathbf{\Psi}_{\mathcal{M}}^{T} \mathbf{y}}{\|\mathbf{y}\|_{2}} \right\|_{\mathcal{A}}^{*} \le \kappa \implies
\frac{1}{\|\mathbf{y}\|_{2}}\left\|\mathbf{\Psi}_{\mathcal{M}}^{T}\mathbf{y}\right\|_{\mathcal{A}}^{*} \le \kappa.
\end{align}
The optimality condition for $\hat{\mathbf{x}}= \mathbf{0}$ for a given $\mathbf{y}\neq\mathbf{0}$ under the ANM framework is given by $\left\|\mathbf{\Psi}_{\mathcal{M}}^{T}\mathbf{y}\right\|_{\mathcal{A}}^{*} \le \kappa$.$\square$

\par We use the Alternating Direction Method of Multipliers (ADMM) algorithm to obtain closed-form updates to the following equivalent SDP formulation of the SR-ANM problem in \eqref{sranm_obj}
\begin{align}
&\min_{\mathbf{T}, \mathbf{x}, t} \| \mathbf{y} - \mathbf{\Psi}_{\mathcal{M}}\mathbf{x}\|_{2} + \frac{\kappa}{2} \left(t+\frac{\operatorname{tr}(\mathbf{T})}{\tilde{M}}\right), \nonumber \\
	& s.t.  \quad \begin{bmatrix}\mathbf{T}  & \mathbf{x}\\ \mathbf{x}^H & t\end{bmatrix}\succeq 0, \quad \mathbf{T} \in \mathcal{T}^{H}.
\label{sranm_obj_sdp}
\end{align}
Defining the auxiliary variable $\mathbf{Q} \in \mathbb{C}^{(\tilde{M}+1)\times(\tilde{M}+1)}$
\begin{align}
\mathbf{Q} = \begin{bmatrix}\mathbf{Q}_{11}  & \mathbf{Q}_{12}\\ \mathbf{Q}_{21} & \mathbf{Q}_{22}\end{bmatrix}, 
\label{q_matrix}
\end{align}
where $\mathbf{Q}_{11} \in \mathbb{C}^{ \tilde{M} \times \tilde{M}}$. The optimization problem in \eqref{sranm_obj_sdp} can be expressed in the following equivalent form where the Toeplitz-Hermitian constraint is decoupled from the PSD constraint
\begin{align}
&\min_{\mathbf{T}, \mathbf{Q}, \mathbf{x}, t} \| \mathbf{y} - \mathbf{\Psi}_{\mathcal{M}}\mathbf{x}\|_{2} + \frac{\kappa}{2} \left(t+\frac{\operatorname{tr}(\mathbf{T})}{\tilde{M}}\right), \nonumber \\
	& s.t.  \quad \mathbf{Q} = \begin{bmatrix}\mathbf{T}  & \mathbf{x}\\ \mathbf{x}^H & t\end{bmatrix}, \quad \mathbf{T} \in \mathcal{T}^{H}, \quad \mathbf{Q}\succeq 0.
\label{sranm_obj_sdp_2}
\end{align}
The augmented Lagrangian is then given by
\begin{align}
L_{\rho}(\mathbf{T},\mathbf{Q},\mathbf{U}, \mathbf{x}, t) = &\| \mathbf{y} - \mathbf{\Psi}_{\mathcal{M}}\mathbf{x}\|_{2} + \frac{\kappa}{2} \left(t+\frac{\operatorname{tr}(\mathbf{T})}{\tilde{M}}\right) \nonumber\\ &+\frac{\rho}{2} \left\| \mathbf{Q}-\begin{bmatrix}\mathbf{T}  & \mathbf{x}\\ \mathbf{x}^H & t\end{bmatrix}+\mathbf{U} \right\|_{F}^{2},
\end{align}
where $\mathbf{U} \in \mathbb{C}^{(\tilde{M}+1)\times(\tilde{M}+1)}$ is the dual variable that follows the same block partitioning as \eqref{q_matrix} and $\rho>0$. The update steps are then given by
\begin{align}
&(\mathbf{T}^{(p+1)}, \mathbf{x}^{(p+1)}, t^{(p+1)}) = \argmin_{\mathbf{T}, \mathbf{x}, t}{L_{\rho}(\mathbf{T},\mathbf{Q}^{(p)},\mathbf{U}^{(p)}, \mathbf{x}, t)}, \label{x_T_t_update}\\
&\mathbf{Q}^{(p+1)} = \argmin_{\mathbf{Q}}{L_{\rho}(\mathbf{T}^{(p+1)},\mathbf{Q},\mathbf{U}^{(p)}, \mathbf{x}^{(p+1)}, t^{(p+1)})}\label{Q_update},\\
&\mathbf{U}^{(p+1)} = \mathbf{U}^{(p)} + \mathbf{Q}^{(p+1)} - \begin{bmatrix}\mathbf{T}^{(p+1)}  &  \mathbf{x}^{(p+1)}\\  (\mathbf{x}^{(p+1)})^H & t^{(p+1)}\end{bmatrix}.\label{U_update}
\end{align}
The update steps for $(\mathbf{T}^{(p)},\mathbf{Q}^{(p)}, t^{(p)})$ follow the same update rules as in classical ANM and are given by
\begin{align}
\mathbf{T}^{(p+1)} &= \mathcal{P}_{\mathcal{T}^{H}}(\mathbf{Q}_{11}^{(p)}+\mathbf{U}_{11}^{(p)}-\frac{\kappa}{2\rho \tilde{M}}\mathbf{I}_{\tilde{M}}), \\
t^{(p+1)} &= \mathbf{Q}_{22}^{(p)}+\mathbf{U}_{22}^{(p)}-\frac{\kappa}{2\rho}, \\
\mathbf{Q}^{(p+1)} &= \mathcal{P}_{\succeq 0}\big( \begin{bmatrix}\mathbf{T}^{(p+1)}  &  \mathbf{x}^{(p+1)}\\  (\mathbf{x}^{(p+1)})^H & t^{(p+1)}\end{bmatrix} - \mathbf{U}^{(p)}\big) \label{psd_anm},
\end{align}
where $\mathcal{P}_{\mathcal{T}^{H}}(\cdot)$ and $\mathcal{P}_{\succeq 0}(\cdot)$ denote the projection operators onto the set of Toeplitz-Hermitian matrices and the set of positive semidefinite (PSD) matrices, respectively, and where $\mathbf{U}$ follows the same block partitioning as $\mathbf{Q}$ in \eqref{q_matrix}. The $\mathbf{x}$-update can be found by minimizing the objective
\begin{align}
\mathbf{x}^{(p+1)} =  \argmin_{\mathbf{x}} {\| \mathbf{y} - \mathbf{\Psi}_{\mathcal{M}}\mathbf{x}\|_{2} + \rho\| \mathbf{x} - \mathbf{v}^{(p)}\|_{2}^{2}}, \label{x_update_2}
\end{align}
where $\mathbf{v}^{(p)} = \mathbf{Q}_{12}^{(p)} + \mathbf{U}_{12}^{(p)}$, and where we made use of the Hermitian symmetry of both $\mathbf{U}$ and $\mathbf{Q}$. Denoting the complement set of $\mathcal{M}$ by $\mathcal{M}_{C} = \{m\in\{1,\ldots,\tilde M\}\mid m\notin\mathcal{M}\}$, we can write \eqref{x_update_2} as the sum of two independent convex optimization problems
\begin{align}
\mathbf{x}^{(p+1)} &=  \argmin_{\mathbf{x}} {}\|\mathbf{y} - \mathbf{x}_{\mathcal{M}}\|_{2} + \rho\| \mathbf{x}_{\mathcal{M}} - \mathbf{v}_{\mathcal{M}}^{(p)}\|_{2}^{2} \nonumber\\
 &+ \rho\| \mathbf{x}_{\mathcal{M}_{C} }- \mathbf{v}_{\mathcal{M}_{C}}^{(p)}\|_{2}^{2}, \nonumber\\
&\textrm{s.t.}\quad\mathbf{x}_{\mathcal{M}} = \mathbf{\Psi}_{\mathcal{M}}\mathbf{x}, \quad \mathbf{x}_{\mathcal{M}_{C}} = \mathbf{\Psi}_{\mathcal{M}_{C}}\mathbf{x}, \label{x_update_3}
\end{align}
where $\mathbf{\Psi}_{\mathcal{M}_{C}} \in \{0,1\}^{(\tilde{M}-M)\times \tilde{M}}$ is the subsampling matrix corresponding to the set of indices $\mathcal{M}_{C}$. The update step for $\mathbf{x}_{\mathcal{M}_{C}}$ is then given by
\begin{align}
\mathbf{x}_{\mathcal{M}_{C}}^{(p+1)} =  \mathbf{v}_{\mathcal{M}_{C}}^{(p)}.
\end{align}
Using the definition of the proximal operator of the $L_{2}$-norm, the update step for $\mathbf{x}_{\mathcal{M}}$ is additionally given by
\begin{align}
\mathbf{x}_{\mathcal{M}}^{(p+1)} = \mathbf{y} + \max{(1-\frac{1}{2\rho\|\mathbf{v}_{\mathcal{M}}^{(p)}-\mathbf{y}\|_{2}},0)}(\mathbf{v}_{\mathcal{M}}^{(p)}-\mathbf{y}).
\end{align}
It is worth noting here that SR-ANM has a computational complexity of $\mathcal{O}(\tilde{M}^{3})$, which is identical to ANM, and is mostly dominated by the PSD projection step in \eqref{psd_anm}.
\section{Experimental Results}
\label{sec:experimental_results}
\subsection{Experimental Setup}
In this section, we perform an experimental verification of the proposed SR-ANM framework and compare its performance against ANM across different SNR levels and regularization levels $\kappa$. We first subsample an array of $M=20$ elements from a ULA with $\tilde{M}=31$ elements, which corresponds to an aperture of $30\lambda/2$. Next, we generate $K=5$ random angles $\{\theta_{k}^{*}\}_{k=1}^{K}$ from $\mathcal{U}(-90^{\circ}, 90^{\circ})$ such that the corresponding normalized frequencies $\{f_{k}^{*}\}_{k=1}^{K}$ with $f_{k}^{*} = -\frac{1}{2} \sin(\theta_{k}^{*})$ have a minimum separation of $1/M$ \cite{supres}. Next, for each source, we generate a complex amplitude $c_{k}=|c_{k}|e^{j\phi_{k}}$ with $|c_{k}|\sim \mathcal{U}(0.05, 1)$ and $\phi_{k} \sim \mathcal{U}(0, 2\pi)$. Finally, we generate $\sigma$ from a given SNR level in dB according to 
\begin{align}
\sigma=\frac{1}{
\sqrt{M}}10^{-\frac{\textrm{SNR}}{20}}\Bigl\|\sum_{k=1}^{K}c_{k}\mathbf{\Psi}_{\mathcal{M}}\mathbf{a}(\theta_{k}^{*})\Bigr\|_{2},
\end{align}
and the final measurement vector is obtained using \eqref{signal_model}. 

\par For the $r$-th measurement vector $\mathbf{y}_{r}$, with $r=1,\ldots, R$ where $R$ is the number of Monte Carlo trials, we first compute the Toeplitz-Hermitian matrix $\hat{\mathbf{T}}_{r}$, using either ANM or SR-ANM, and subsequently we recover the set of angles $\mathcal{S}_{r}=\{\hat{\theta}_{k,r}\}_{k=1}^{K}$ by performing a Root-MUSIC estimation on $\hat{\mathbf{T}}_{r}$ where the model order $K$ is directly provided for both ANM and SR-ANM. A ground truth angle $\theta_{k,r}^{*}$ is considered successfully recovered if its estimate $\hat{\theta}_{k,r}$ is within a maximum angular distance of $\Delta\theta = 10^{\circ}$,
i.e. $\min (|\hat{\theta}_{k,r}-\theta_{k,r}^{*}|, 180^{\circ}-|\hat{\theta}_{k,r}-\theta_{k,r}^{*}|) \le 10^{\circ}$. Denoting by $\mathcal{I}_{d,r}\subseteq \{1,\ldots,K\}$ the subset of indices at which a positive detection occurred,  the number of detections $T_{d,r}$ for the $r$-th trial is then simply given by $T_{d,r} = |\mathcal{I}_{d,r}|$, where $|\mathcal{Q}|$ denotes the cardinality of the set $\mathcal{Q}$. The detection rate $P_{d}$ is then given by 
\begin{align}
P_{d} = \frac{1}{R}\sum_{r=1}^{R}\frac{T_{d,r}}{K}.
\end{align} 
The angular root-mean-squared error (RMSE) is additionally computed for the trials where at least one true detection occurred, i.e. $T_{d,r} \neq 0$, as follows
\begin{align}
\textrm{RMSE} = \sqrt{\frac{1}{R_{d}}\sum_{\substack{r=1 \\ T_{d,r}\neq 0}}^{R}\frac{1}{T_{d,r}}\sum_{k \in \mathcal{I}_{d,r}} (\theta_{k,r}^{*}-\hat{\theta}_{k,r})^{2}},
\end{align}
where $R_{d} = \sum_{r=1}^{R}\mathds{1}(T_{d,r}\neq 0)$.
\par For ANM, we set the regularizer to $\kappa = \beta \sigma(M\log\tilde{M})^{1/2}$ where we directly provide the noise standard-deviation $\sigma$ for each trial and where $\beta \in \{0.5, 1, 1.5\}$. For SR-ANM, we set $\kappa = \beta (\log\tilde{M})^{1/2}$. The SNR level is varied from 0 dB to 40 dB with a stepsize of 5 dB. At each SNR level, we perform $R=1000$ Monte Carlo trials. We use a stopping criterion given by $\frac{\|\mathbf{t}^{(p+1)}-\mathbf{t}^{(p)}\|_{2}}{\|\mathbf{t}^{(p)}\|_{2}}<10^{-4}$ where $\mathbf{t}^{(p)}$ is the vector parametrizing $\mathbf{T}^{(p)}$, i.e. the first column of $\mathbf{T}^{(p)}$. The maximum number of ADMM iterations is set to $p_{max} = 5000$. For both ANM and SR-ANM, we vary the ADMM parameter $\rho\in\{0.1, 1, 10\}$. We initialize all the vectors and matrices in the ADMM implementation of both ANM and SR-ANM with null vectors and matrices, respectively.
\begin{figure}[!t]
    \centering
    \begin{subfigure}[t]{\columnwidth}
        \centering
        \includegraphics[width=6.5cm, height = 4.5cm]{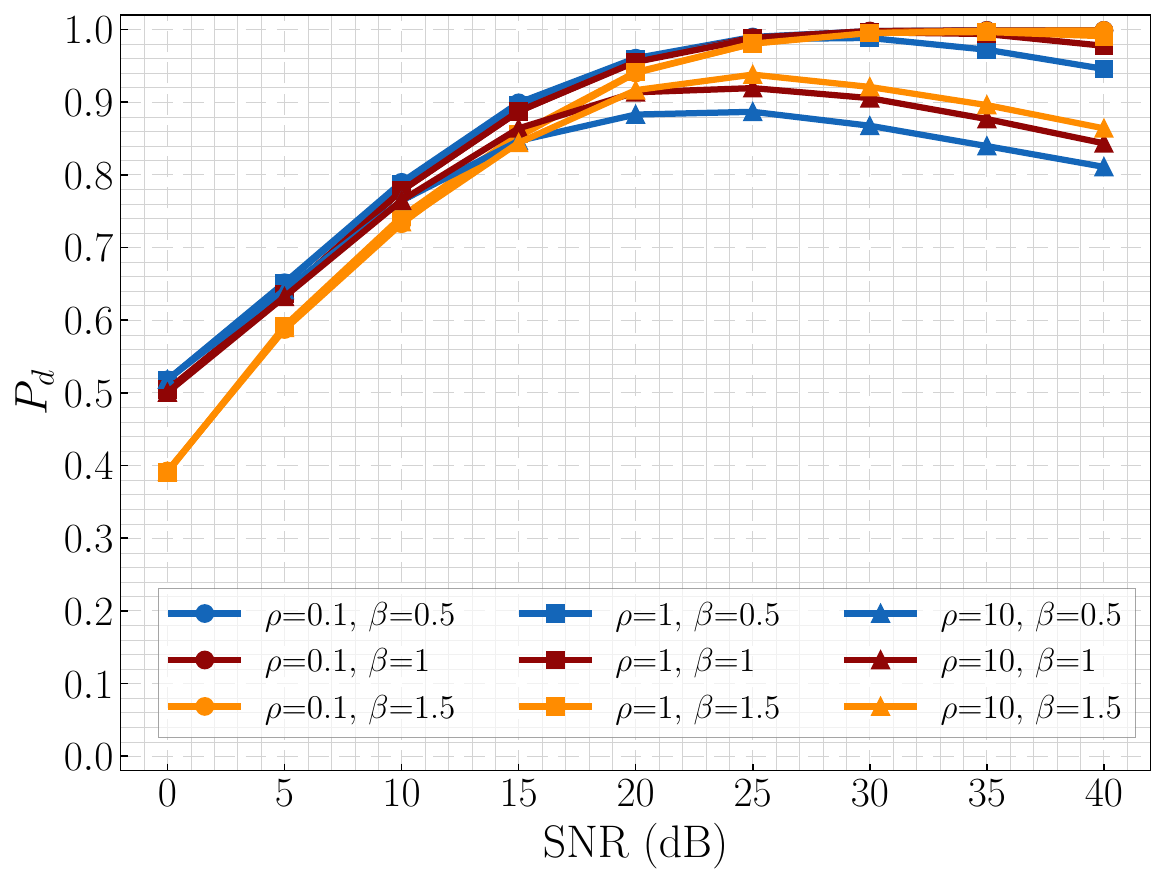}
        \caption{$P_{d}$ (ANM)}
        \label{anm_detection_rate}
    \end{subfigure}
    \par\medskip
    \begin{subfigure}[t]{\columnwidth}
        \centering
        \includegraphics[width=6.5cm, height = 4.5cm]{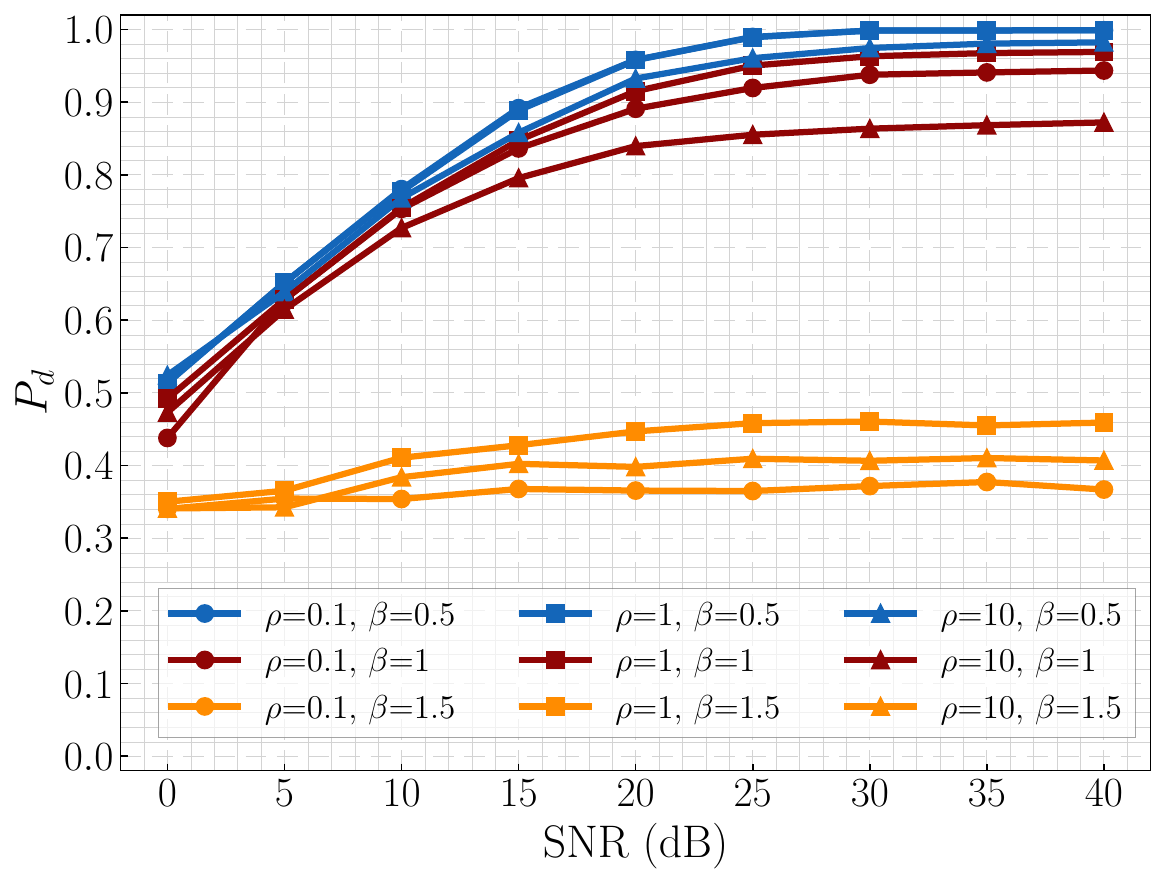}
        \caption{$P_{d}$ (SR-ANM)}
        \label{sr_anm_detection_rate}
    \end{subfigure}
    \caption{Detection rate $P_{d}$ as a function of the atomic norm regularization parameter $\kappa$, the ADMM hyperparameter $\rho$, and the SNR for ANM (a) and SR-ANM (b).}
    \label{fig:detection_rate}
\end{figure}
\begin{figure}[!t]
    \centering
    \begin{subfigure}[t]{\columnwidth}
        \centering
        \includegraphics[width=6.5cm, height = 4.5cm]{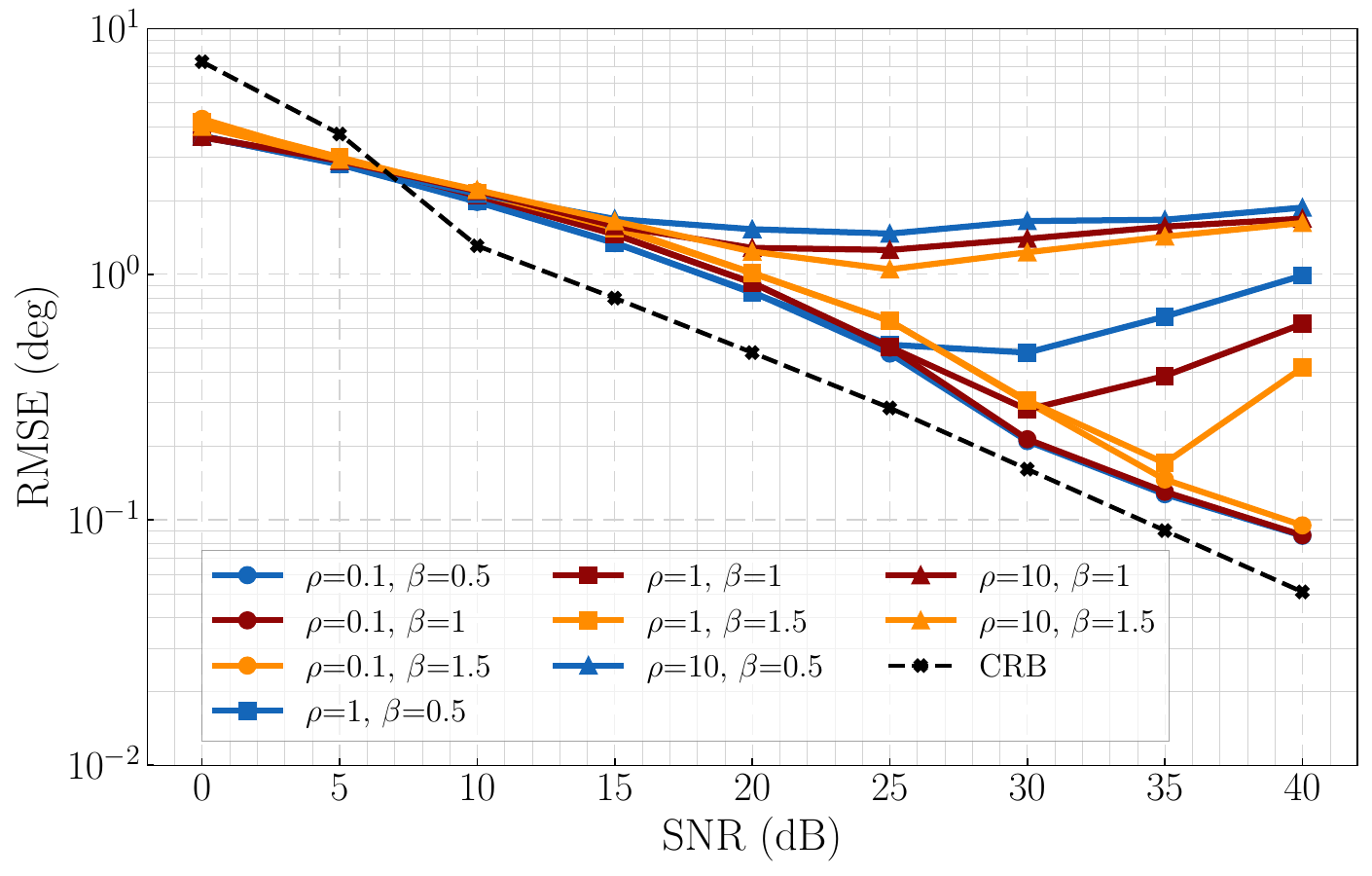}
        \caption{RMSE (ANM)}
        \label{anm_rmse}
    \end{subfigure}
    \par\medskip
    \begin{subfigure}[t]{\columnwidth}
        \centering
        \includegraphics[width=6.5cm, height = 4.5cm]{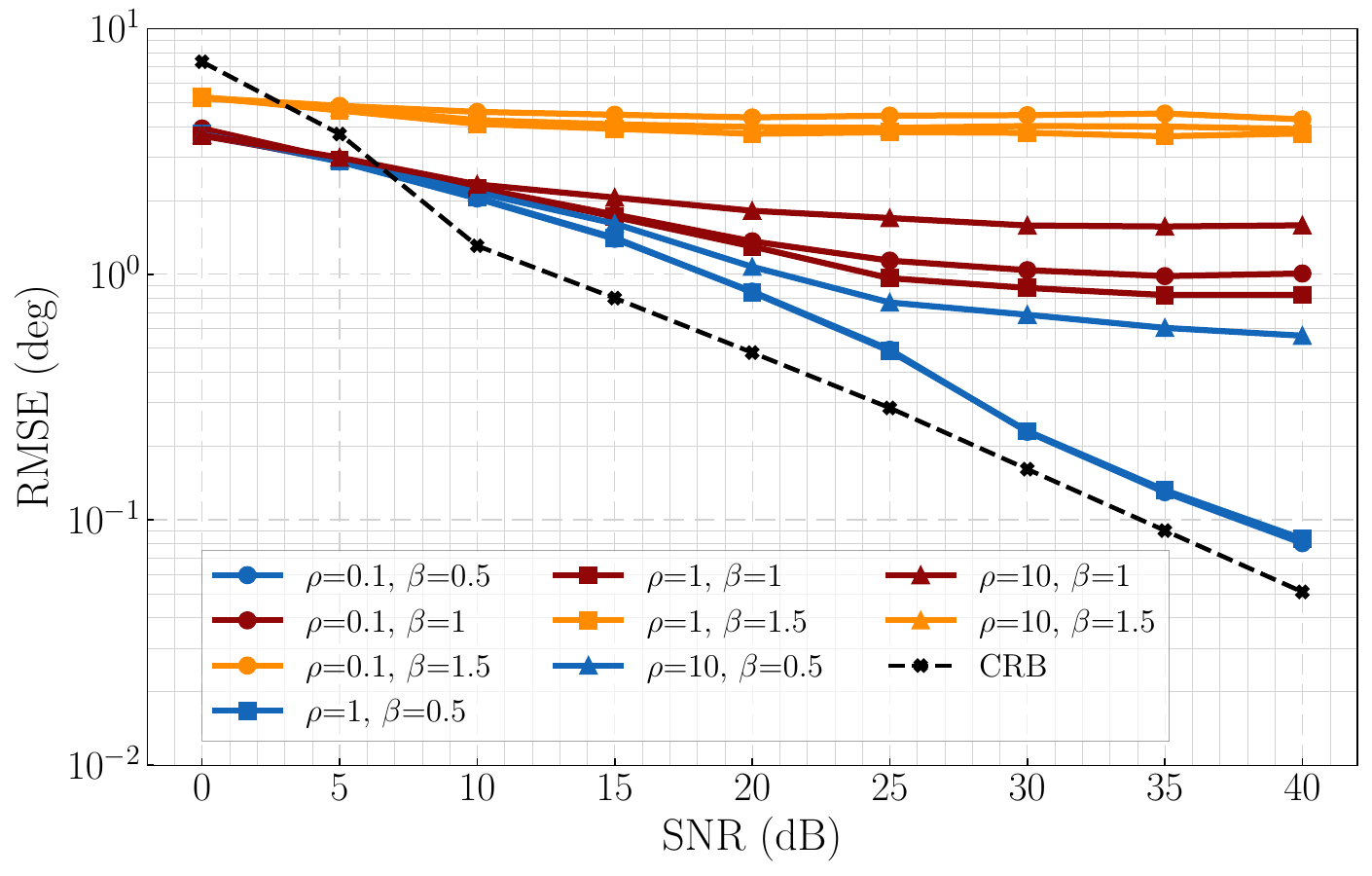}
        \caption{RMSE (SR-ANM)}
        \label{sr_anm_rmse}
    \end{subfigure}
    \caption{Angular RMSE as a function of the atomic norm regularization parameter $\kappa$, the ADMM hyperparameter $\rho$, and the SNR for ANM (a) and SR-ANM (b).}
    \label{fig:rmse}
\end{figure}
\subsection{Results Discussion}
As can be seen in Fig. \ref{anm_detection_rate} and Fig. \ref{anm_rmse}, the ANM estimator is asymptotically efficient for $\rho = 0.1$ and $\beta = 1$, which corresponds to $\kappa = \sigma(M\log\tilde{M})^{1/2}$. We can also see that ANM is able to achieve the same efficiency with $\beta = 0.5$, which indicates that the minimum regularizer value required to guarantee the asymptotic efficiency, namely $\kappa = \sigma(M\log\tilde{M})^{1/2}$,  is sufficient but not necessary. We can additionally see from Fig. \ref{sr_anm_detection_rate} and Fig. \ref{sr_anm_rmse} that the SR-ANM estimator is also able to achieve the same asymptotic efficiency for $(\rho,\beta) = (1, 0.5)$ along with $(\rho, \beta) = (0.1, 0.5)$, which corresponds to $\kappa = (\log\tilde{M})^{1/2}/2$, i.e. without requiring prior knowledge of $\sigma$.

\section{Conclusion}
\label{sec:conclusion}
This paper proposed a gridless sparse recovery estimator for DoA estimation, termed SR-ANM, which combines the atomic norm regularization with an $L_{2}$-norm residual similar to the Square-root LASSO estimator so as to obtain increased robustness of the atomic norm regularization parameter to the noise level. Experimental verification of the proposed method confirmed the estimator is asymptotically efficient without requiring prior knowledge of the noise standard deviation.
\bibliographystyle{IEEEtran}
\bibliography{refs}

\end{document}